\documentstyle[aps,prl,multicol,eqsecnum]{revtex}
 \renewcommand{\narrowtext}{\begin{multicols}{2} \global\columnwidth20.5pc}
 \renewcommand{\widetext}{\end{multicols} \global\columnwidth42.5pc}  

\input{epsf.tex}

\def\top#1{\vskip #1\begin{picture}(290,80)(80,500)\thinlines \put(
65,500){\line( 1, 0){255}}\put(320,500){\line( 0, 1){
5}}\end{picture}}
\def\bottom#1{\vskip #1\begin{picture}(290,80)(80,500)\thinlines \put(
330,500){\line( 1, 0){255}}\put(330,500){\line( 0, -1){
5}}\end{picture}}

\begin{document}
\title{Chern-Simons versus dipolar composite 
fermions at finite wavevector}

\author{Yong Baek Kim}
\address{
Department of Physics, The Ohio State University, 
Columbus, OH 43210}

\date{\today}
\maketitle

\begin{abstract}
It was recently shown that dipolar composite fermions 
emerged from the lowest-Landau-level formulation of the 
quantum Hall effect give rise to similar
results as those of the Chern-Simons gauge
theory in the long wavelength and low energy limit. 
We ask whether this correspondence is still valid 
at finite wavevectors where the excitations do not
necessarily look like dipolar quasiparticles.
In particular, $q=2k_F$ density-density response 
function of the compressible state at $\nu=1/2$ is 
evaluated in the low energy limit within the framework
of the lowest-Landau-level theory.
The imaginary parts of the density-density response functions
at $q=2k_F$ of two theories have the same $\sqrt{\omega}$
dependence. However, the coefficient of the $\sqrt{\omega}$
term in the case of the lowest-Landau-level theory is not
universal and can be much smaller than the corresponding
coefficient in the Chern-Simons theory.
We also discuss
possible connection between these results and the
recent experiment on phonon-mediated drag in the double-layer
$\nu=1/2$ system. 

\end{abstract}


\narrowtext

\section{Introduction} 

The compressible state of two dimensional
electrons at the filling factor $\nu=1/2$ has
been explained as the result of Fermi surface
formation of quasiparticles called composite 
fermions \cite{hal,stormer}.
Here the filling factor, $\nu$, is the ratio
of the number of electrons and the total number
of external flux quanta given by the external
magnetic field. 
Thus there exist ${\tilde \phi}$ 
number of external flux quanta per electron if
$\nu = 1/{\tilde \phi}$.    
The composite fermions are bound objects of electrons
and their correlation holes. Due to the fact that 
the electron wavefunction vanishes at the positions 
of correlation holes and there exists a phase winding of 
the electron wavefunction around them,
these correlation holes are also called vortices. 
The bound state of electrons and even number of vortices 
satisfies Fermi statistics leading to the name 
``composite fermions'' \cite{jain}.

Upon the discovery of the compressible states in the
half-filled Landau level, it was soon proposed that
the vortices can be represented by fictitious flux
quanta pointing in the opposite direction of the
external magnetic field \cite{HLR}. 
In this case, the bound state 
of an electron and two fictitious flux quanta would 
see zero effective magnetic field on average 
at $\nu=1/2$ \cite{HLR,zhang}.
This is because the number of external flux quanta
per electron and that of fictitious flux quanta 
are the same. Now the composite fermion system can form
a Fermi sea as if they are free fermions in zero magnetic
field. These ideas can be formulated in terms of the Chern-Simons
gauge theory that represents the fictitious flux quanta
by a U(1) gauge field and requires the position of these
flux quanta to be the same as those of electrons. 
This ``Chern-Simons composite fermion'' theory explained 
the finite compressibility of the system and various 
experimental results that require understanding of the
long wavelength and low energy properties of the 
system \cite{HLR}.

In spite of this success, the Chern-Simons composite fermion
theory suffers from some important problems.
At the most naive level, the flux attachment transformation 
uses the kinetic energy of the electrons to generate the 
kinetic energy of composite fermions. If this transformation 
is taken seriously, the mass of the composite fermions is given by
the bare mass of electrons. This is certainly not correct
because, in the lowest Landau level, the kinetic energy of 
electrons is quenched and the only relevant energy scale is 
given by the interaction energy; the mass of the composite 
fermions should be determined by the interaction 
energy scale \cite{hal,HLR}. 
Therefore the kinetic energy term with the bare mass in the 
Hamiltonian should be somehow replaced by the term with the 
interaction induced mass \cite{hal,HLR}.  
In the Chern-Simons theory approach, the kinetic energy 
term with the interaction induced mass has been used on 
phenomenological ground \cite{hal,HLR}. 
One hopes that the final answer
will be correct in the limit of the zero bare mass
(infinite cyclotron energy).  

Sometime ago, more microscopic picture was suggested to resolve
the unsatisfactory aspects of the Chern-Simons 
theory \cite{read94}.
In the lowest Landau level, the system of $N$ number of 
composite fermions maintains the antisymmetry of the total 
wavefunction by assigning a different vector 
${\bf k}_j$ ($j=1,...,N$)
to each composite fermion such that the distance between 
the electron and vortices in a given composite fermion $j$ is 
$|{\bf k}_j| l^2_B$ and the direction of ${\bf k}_j$ is 
perpendicular to the line connecting the positions of 
electron and vortices \cite{read94}.
Here $l_B = \sqrt{\hbar c/eB}$ is the magnetic length. 
The electron and vortices within each composite fermion drift 
along the equipotential, $V(|{\bf k}|)$, in the same direction 
keeping the distance, $|{\bf k}| l^2_B$, between them. 
This can generate the dispersion (or ``kinetic energy'') of 
the composite fermions \cite{read94}. 
Also the vectors ${\bf k}_j$ form a Fermi sea in 
${\bf k}$-space with a well defined Fermi wavevector 
$k_F = \sqrt{4 \pi n_e}$ where $n_e$ is the density of
electrons. Since the ``kinetic energy'' arises from
the interaction potential, the effective mass at the
Fermi wavevector is determined by the interaction 
energy scale \cite{read94}.        
These composite fermions are called ``dipolar composite
fermions'' due to the dipolar internal 
structure \cite{read94,shankar,dhlee,pasquier,stern,read98,shankar99}.

The ideas outlined above have been recently formulated 
in several different 
ways \cite{shankar,dhlee,pasquier,stern,read98,shankar99}. 
The essential parts of the 
formulation can be summarized as 
follows \cite{shankar,dhlee,pasquier,stern,read98,shankar99}. 
We consider
the case of $\nu = 1/{\tilde \phi}$ for simplicity. 
Let ${\bf R}_{ej}$ be the position of the electron $j$
and ${\bf R}_{vj}$ be the position of ${\tilde \phi}$ 
number of vortices associated with the electron $j$.
Equivalently one can regard the complex coordinates of 
vortices, $w_j$, as dummy variables in reference to the
coordinates, $z_j = x_j + i y_j$, of the electrons.  
Using a canonical transformation, we can define 
an equivalent set of coordinates:
\begin{equation}
{\bf r}_j = {1 \over 2} ({\bf R}_{ej} + {\bf R}_{vj}) \ , 
\ \ \  
{\bf k}_j = - \wedge ({\bf R}_{ej} - {\bf R}_{vj}) l^{-2}_B\ ,
\end{equation}
where $(\wedge {\bf a})_{\mu} \equiv 
\epsilon_{\mu \nu} a_{\nu}$ ($\mu, \nu = x,y$)
and ${\bf a} \wedge {\bf b} = {\bf a} \cdot (\wedge {\bf b})$.
In this representation, ${\bf r}_j$ is the center of
mass coordinate, $|{\bf k}_j|l^2_B$ is the distance between
the electron and vortices, and the direction of ${\bf k}_j$
is perpendicular to the line connecting the electron and
the vortices.  
Notice that ${\bf R}_{ej}$ and ${\bf R}_{vj}$ are
guiding center coordinates of electrons and vortices 
so that they satisfy the following commutation relations.
\begin{equation}
[R^x_{ej}, R^y_{ej}] = -il^2_B \ ,  
\ \ \ 
[R^x_{vj}, R^y_{vj}] = il^2_B \ , 
\end{equation}
where $R^x_{ej}$ and $R^y_{ej}$ denote the $x$ and $y$ components
of ${\bf R}_{ej}$ and so on.
Using these relations, we get 
\begin{equation}
[r^{\alpha}_j, k^{\beta}_j] = i \delta_{\alpha \beta} \ . 
\end{equation}
where $\alpha, \beta = x, y$.
Thus ${\bf k}_j$ serves as the canonical momentum of the 
coordinate ${\bf r}_j$.
The density operators for electrons and vortices can be
written in the Fourier space as 
\begin{eqnarray}
\rho^L ({\bf q}) = \sum_j e^{i {\bf q} \cdot {\bf R}_{ej}}
= \sum_j e^{i {\bf q} \cdot [{\bf r}_j - 
\wedge {\bf k}_j (l^2_B/2)]} \ , \cr 
\rho^R ({\bf q}) = \sum_j e^{i {\bf q} \cdot {\bf R}_{vj}}
= \sum_j e^{i {\bf q} \cdot [{\bf r}_j + 
\wedge {\bf k}_j (l^2_B/2) ] } \ ,
\label{density} 
\end{eqnarray}
where $L$ and $R$ represent the electron and vortices
respectively. In order to eliminate additional degrees
of freedom (the coordinates of the vortices are dummy
variables) and ensure the equivalence to the original
system, we require the following constraint.
\begin{equation}
\rho^R ({\bf q}) = {\bar \rho} (2 \pi)^2 
\delta ({\bf q}) \ ,
\label{const}
\end{equation} 
where ${\bar \rho}$ is the average density of electrons.
The Hamiltonian of the system is written in terms of
physical electron density operators 
\cite{shankar,pasquier,read98}:
\begin{equation}
H = {1 \over 2} \int {d^2k \over (2 \pi)^2}
{\tilde V}({\bf q}) 
: \rho^L ({\bf q}) \rho^L (-{\bf q}) : \ ,
\label{ham}
\end{equation}
where ${\tilde V}({\bf q}) = V({\bf q}) e^{-q^2 l_B^2 / 2}$ 
and $V({\bf q})$ is the interaction potential.
The colons $:...:$ represents the normal ordering.

In the long wavelength limit, the electron density 
operator can be expanded as
\begin{equation}
\rho^L ({\bf q}) \approx 
{\bar \rho} (2 \pi)^2 
\delta ({\bf q}) - i {l^2_B \over 2} 
\sum_j e^{i {\bf q} \cdot {\bf r}_j} 
{\bf q} \wedge {\bf k}_j \ . 
\end{equation}
The Hamiltonian can now be written as (up to a constant)
\begin{equation}
H = {l^2_B \over 8} \sum_{ij} \int 
{d^2k \over (2 \pi)^2}
{\tilde V}({\bf q}) ({\bf q} \wedge {\bf k}_i)
({\bf q} \wedge {\bf k}_j) 
e^{i {\bf q} \cdot {\bf r}_i}
e^{-i {\bf q} \cdot {\bf r}_j} \ .
\end{equation}
In particular, the $i=j$ term will give rise to
\begin{equation}
H \sim \sum_j {{\bf k}_j^2 / 2 m^*} \ \
{\rm with} \ \ 
{1 \over m^*} \propto \int {d^2k \over (2 \pi)^2}
{\tilde V}({\bf q}) q^2 l^2_B\ . 
\label{appH}
\end{equation}
This may be taken as the origin of the kinetic 
energy term of the composite fermions in the long wavelength 
limit \cite{shankar,dhlee,pasquier,stern,read98,shankar99}.

It turns out that it is very important to 
incorporate the residual interaction terms and 
the constraint to satisfy proper conservation laws.
For example, the original system of dipolar composite
fermions is invariant under ${\bf k}_j
\rightarrow {\bf k}_j + {\bf K}$ for all $j$ and
any ${\bf K}$ because it just corresponds to a 
constant shift in the center of mass coordinate of the entire 
system \cite{pasquier,stern,read98}. 
This symmetry is intimately related to
the fact that the kinetic energy term of the electrons
is absent or the system is in the infinite bare mass
limit $m_b \rightarrow \infty$. 
This, for example, leads to $F_1 = -1$ among
Landau parameters via $m^*/m_b = 1 + F_1$ if the 
dipolar composite fermions are treated as quasiparticles 
in the Landau Fermi liquid theory \cite{stern,read98}. 
The approximate Hamiltonian in Eq.\ref{appH}, however,
breaks this symmetry. 
While this problem can be fixed in the long wavelength
and low energy limits by incorporating the
constraints \cite{stern}, 
it was also shown that one can formulate
the theory in a conserving approximation for all
${\bf q}$ and $\omega$ (more specifically for 
$\nu=1$ bosons) \cite{read98}. 
In any case, the conserving approximation leads to
the conclusion that the system is still 
compressible \cite{stern,read98}.
Also various physical response functions in the low
energy and long wavelength limit were shown to 
have the same forms as those of the Chern-Simons 
gauge theory approach. In other words, the dipolar 
composite fermion description in the infinite bare mass
limit seems to be equivalent 
to the Chern-Simons 
composite fermion system in the zero bare mass 
(or infinite cyclotron energy) limit
as far as the low energy and long wavelength limits
are concerned \cite{stern,read98}.

In this paper, we ask whether the close correspondence 
between the results of the Chern-Simons theory and 
those of the lowest Landau level theory is 
still valid at finite wavevectors where the excitations 
do not necessarily look like dipolar quasiparticles.
We were partly motivated by the recent phonon drag 
experiment in the double-layer $\nu=1/2$ 
system \cite{gramila99}.
In this experiment, the drag resistivity between two
layers due to the electron-phonon interaction 
is measured \cite{gramila99}. 
In the absence of an applied magnetic field, it has been known
that the phonon drag divided by $T^2$ reaches its maximum when 
the temperature becomes of the order of $T_0 = c (2k^e_F)$
where $c$ is the phonon velocity and $k^e_F$ is the
Fermi wavevector of the electrons in each 
layer \cite{gramila91,gramila93}.
This is due to the fact that the particle-hole excitations
of electrons cease to exist beyond the wavevector 
$q=2k^e_F$ so that the phonons with $q > 2k^e_F$ cannot scatter
electrons at low temperatures. 
Thus the scattering between the particle-hole 
continuum and the phonons is suppressed when the energy
scale is larger than $T_0$. 
In a way, this experiment can tell us the properties of 
the system at short distances like $(k^e_F)^{-1}$.
In the case of $\nu=1/2$ double-layer system, one may expect
that the cutoff wavevector scale would be set by the
composite fermion Fermi wavevector, 
$k^{\rm cf}_F$ where $k^{\rm cf}_F = \sqrt{2} k^e_F$ if
all the spins of electrons are polarized due to strong
magnetic field. It is expected that the maximum 
of the drag resistivity should occur at a temperature
$T^{\rm max}_{1/2}$ around 
$T_{1/2} = c (2 k^{\rm cf}_F) = \sqrt{2} T_0$. 
On the other hand, in the experiment, $T^{\rm max}_{1/2}$ 
turns out to be even smaller than $T_0$ \cite{gramila99}.
Recently, Bonsager, MacDonald, and the author 
performed a theoretical calculation of the drag resistivity
in the Chern-Simons theory \cite{bonsager}. 
It was found that the maximum
of the drag resistivity indeed occurs around $T_{1/2}$ 
if the effective Fermi energy (determined from the 
effective mass) of the composite fermions 
$\varepsilon_{F}$ is much larger than 
$T_{1/2}$ \cite{bonsager}.
However, for realistic values of effective mass, 
one finds $\varepsilon_F \sim T_{1/2}$ leading to 
substantial finite temperature effects that were
not observed in the experiment \cite{bonsager}. 
At this stage, it appears 
that the Chern-Simons theory does not capture the correct 
short distance properties of the system.  

Upon this situation, it is natural to ask what happens
to the composite fermions in the lowest-Landau-level 
theory at short distances because this approach is 
supposed to be more microscopic. One may expect that
more microscopic theory may give rise to different
results compared to the predictions of the Chern-Simons
theory and eventually explain the experiment. 
In this paper, we will consider the density-density 
response function at $q=2k_F$
at low energies and investigate possible difference
between the results of the lowest Landau level theory
and the Chern-Simons theory. 
In order to compare the theory with the drag experiment, 
we need to know the physical response functions at 
arbitrary energy and wavevector scales. 
Thus complete explanation of the experiment mentioned above 
is beyond the scope of this paper.
We will, however, try to make
a connection wherever it is possible and discuss what 
might happen in the drag experiment using our results.         

The rest of the paper is organized as follows.
In section II, we briefly review the formalism of 
the lowest-Landau-level theory. In section III, the
density-density response function at $q=2k_F$ in the low energy
limit is evaluated. In section IV, these results
are compared with those of the Chern-Simons theory.
In section V, we discuss possible connection between
our results and the phonon drag experiment.
We summarize our results in section VI.        

\section{brief review of the lowest-Landau-level theory}

We set $l_B=1$ from now on. We restore this factor explicitly
wherever it is necessary. 
It can be shown that the density operators of electrons and vortices 
in Eq.\ref{density} can be written in second quantized 
forms as \cite{read98}
\begin{eqnarray}
\rho^L ({\bf q}) &=& \int {d^2 k \over (2\pi)^2} 
e^{-{1 \over 2} i {\bf k} \wedge {\bf q}} 
c^{\dagger}_{{\bf k}-{1 \over 2}{\bf q}}
c_{{\bf k}+{1 \over 2}{\bf q}}  \ , \cr
\rho^R ({\bf q}) &=& \int {d^2 k \over (2\pi)^2} 
e^{{1 \over 2} i {\bf k} \wedge {\bf q}} 
c^{\dagger}_{{\bf k}-{1 \over 2}{\bf q}}
c_{{\bf k}+{1 \over 2}{\bf q}} \ ,
\end{eqnarray}
where $c_{\bf k}, c^{\dagger}_{\bf k}$ satisfy
the fermionic anticommutation relation;
\begin{equation}
\{ c_{\bf k} , c^{\dagger}_{\bf k'} \} = (2 \pi)^2 
\delta ({\bf k}-{\bf k'}) \ .
\end{equation} 
These operators satisfy the following lowest-Landau-level
algebra first noticed in Ref.\cite{GMP}.
\begin{eqnarray}
[\rho^L ({\bf q}), \rho^L ({\bf q'})] &=& 
- 2 i {\rm sin} \left (
{{\bf q} \wedge {\bf q'} \over 2} \right )
\rho^L ({\bf q}+{\bf q'}) \ , \cr
[\rho^R ({\bf q}), \rho^R ({\bf q'})] &=& 
2 i {\rm sin} \left ( 
{{\bf q} \wedge {\bf q'} \over 2} \right )
\rho^R ({\bf q}+{\bf q'}) \ , \cr
[\rho^L ({\bf q}), \rho^R ({\bf q'})] &=& 0 \ .
\end{eqnarray}
The constraint in Eq.\ref{const} implies that 
\begin{equation}
[\rho^R ({\bf q}) - 
{\bar \rho} (2 \pi)^2 \delta ({\bf q})] 
|\Psi_{\rm phys} \rangle = 0 \ .
\end{equation}
One can first build up states as combinations of
$\prod_{\{ {\bf k}_i \}} 
c^{\dagger}_{\bf k} |0 \rangle$ where $i=1,...,N$
and then project them to satisfy the constraint.
Notice also that the constraint operators 
$G ({\bf q}) = \rho^R ({\bf q}) - 
{\bar \rho} (2 \pi)^2 \delta ({\bf q})$ commutes
with the Hamiltonian given by Eq.\ref{ham}.

Using the second quantized electron density operator,
the Hamiltonian in Eq.\ref{ham} can be 
rewritten as \cite{read98}
\begin{eqnarray}
H &=& {1 \over 2} \int {d^2 k_1 d^2 k_2 d^2 q \over
(2\pi)^6} {\tilde V}({\bf q}) 
e^{{1 \over 2} i {\bf k}_1 \wedge {\bf q} -
{1 \over 2} i {\bf k}_2 \wedge {\bf q}} \cr
&&\times \ c^{\dagger}_{{\bf k}_1 - {1 \over 2}{\bf q}} 
c^{\dagger}_{{\bf k}_2 + {1 \over 2}{\bf q}} 
c_{{\bf k}_2 - {1 \over 2}{\bf q}}
c_{{\bf k}_1 + {1 \over 2}{\bf q}} \ , 
\end{eqnarray}
subject to the constraints 
$\rho^R ({\bf q}) - {\bar \rho} (2\pi)^2 
\delta ({\bf q}) = 0$. 
We first use the Hartree-Fock (HF) approximation 
to obtain the effective kinetic energy of the 
quasiparticles. 
In the HF approximation, the effective single-particle
Hamiltonian can be 
written as \cite{shankar,pasquier,read98,shankar99}
\begin{equation}
H_{\rm eff} = \sum_{\bf k} \xi_{\bf k} 
c^{\dagger}_{\bf k} c_{\bf k} \ ,
\end{equation}
where $\xi_{\bf k} = \varepsilon_{\bf k} - \mu$ and
\begin{equation}
\varepsilon_{\bf k} = {\tilde V}({\bf 0}) 
\int {d^2 k' \over (2\pi)^2} n^0_{\bf k'} - 
\int {d^2 k' \over (2\pi)^2} 
{\tilde V}({\bf k}-{\bf k'}) n^0_{\bf k'} \ . 
\end{equation}
In the ground state at zero temperature, 
$n^0_{\bf k} = \theta (k_F-k)$ with $k_F$
being $\sqrt{2}$ times the Fermi wavevector of
the electrons in zero magnetic field and
$\mu$ chosen such that $\xi_{k_F} = 0$.
Notice that ${\bf k}$ dependence of $\xi_{\bf k}$
comes from the Fock term in the effective Hamiltonian.
When the interaction potential is repulsive, $\xi_{\bf k}$
is a monotonically increasing function of $|{\bf k}|$. 
The effective mass of the quasiparticles at the Fermi
level can be obtained from \cite{read98}
\begin{equation}
{k_F \over m^*} \equiv
{\partial \xi_{\bf k} \over \partial |{\bf k}|}
= - {k_F \over 2 \pi} 
\int {d \theta_{{\bf k}{\bf k'}} \over 2 \pi}
{\tilde V}({\bf k'}-{\bf k}) \
{\rm cos} \ \theta_{{\bf k}{\bf k'}} \ ,
\end{equation}
where $\theta_{{\bf k}{\bf k'}}$ is the angle between
${\bf k}$ and ${\bf k'}$.

The HF ground state which is just the Fermi sea 
$| FS \rangle$ is not annihilated by the constraint
operator; $G ({\bf q}) | FS \rangle \not= 0$ for
${\bf q} \not= 0$. Also $G({\bf q})$ does not commute
with the HF effective Hamiltonian. Thus $G({\bf q})$
are not conserved by the HF approximation.
In order to recover the conserved quantities,
we use the generalized HF theory called
the time-dependent HF approximation which is 
the natural conserving approximation related
to the HF theory.

The density-density response function in this
generalized HF approximation corresponds to 
the sum of all ring and ladder diagrams with
the single particle Green's function given by
\begin{equation}
{\cal G} ({\bf k},\omega_n) = 
(i\omega_n-\xi_{\bf k})^{-1} \ ,
\end{equation} 
where $\omega_n$ is
the Matsubara frequency.
The irreducible density-density response function
of electrons, $\chi^{irr}_{LL}$,
in the generalized HF approximation corresponds
to the sum of the ladder diagrams. The expression
for $\chi^{irr}_{LL}$ was found to have the 
following form \cite{read98}.  

\widetext
\top{-2.8cm}
\begin{eqnarray}
\chi^{irr}_{LL} = &-& \int {d^2 k \over (2\pi)^2}
(e^{i {\bf k} \wedge {\bf q}}-1)
(e^{-i {\bf k} \wedge {\bf q}}-1)
{f(\xi_{{\bf k} + {1 \over 2}{\bf q}})-
f(\xi_{{\bf k} - {1 \over 2}{\bf q}}) \over
\xi_{{\bf k} + {1 \over 2}{\bf q}} - 
\xi_{{\bf k} - {1 \over 2}{\bf q}} - i \omega_{\nu}}
+ \int {d^2 k \ d^2 k' \over (2\pi)^4} 
(e^{i {\bf k} \wedge {\bf q}}-1)
{f(\xi_{{\bf k} + {1 \over 2}{\bf q}})-
f(\xi_{{\bf k} - {1 \over 2}{\bf q}}) \over
\xi_{{\bf k} + {1 \over 2}{\bf q}} - 
\xi_{{\bf k} - {1 \over 2}{\bf q}} - i \omega_{\nu}} \cr
&\times& \Gamma ({\bf k},{\bf k'},{\bf q},i\omega_{\nu})
{f(\xi_{{\bf k'} + {1 \over 2}{\bf q}})-
f(\xi_{{\bf k'} - {1 \over 2}{\bf q}}) \over
\xi_{{\bf k'} + {1 \over 2}{\bf q}} - 
\xi_{{\bf k'} - {1 \over 2}{\bf q}} - i \omega_{\nu}}
(e^{-i {\bf k'} \wedge {\bf q}}-1) \ .
\end{eqnarray}
Here the scattering vertex function, 
$\Gamma ({\bf k},{\bf k'},{\bf q},i\omega_{\nu})$, 
satisfies the following integral equation.
\begin{equation}
\Gamma ({\bf k},{\bf k'},{\bf q},i\omega_{\nu}) = 
{\tilde V}({\bf k}-{\bf k'}) - 
\int {d^2 k_1 \over (2\pi)^2}
\Gamma ({\bf k},{\bf k}_1,{\bf q},i\omega_{\nu})
{f(\xi_{{\bf k}_1 + {1 \over 2}{\bf q}})-
f(\xi_{{\bf k}_1 - {1 \over 2}{\bf q}}) \over
\xi_{{\bf k}_1 + {1 \over 2}{\bf q}} - 
\xi_{{\bf k}_1 - {1 \over 2}{\bf q}} - i \omega_{\nu}}
{\tilde V}({\bf k}_1-{\bf k'}) \ .
\label{integral}
\end{equation}  
It is also useful to define the following one-particle-irreducible
vertex function;
\begin{equation}
\Lambda ({\bf k},{\bf q},i\omega_{\nu})
= 1 - \int {d^2 k_1 \over (2\pi)^2}
{f(\xi_{{\bf k}_1 + {1 \over 2}{\bf q}})-
f(\xi_{{\bf k}_1 - {1 \over 2}{\bf q}}) \over
\xi_{{\bf k}_1 + {1 \over 2}{\bf q}} - 
\xi_{{\bf k}_1 - {1 \over 2}{\bf q}} - i \omega_{\nu}}
\Gamma ({\bf k}_1,{\bf k},{\bf q},i\omega_{\nu}) \ .
\end{equation}
\bottom{-2.7cm}
\narrowtext

One can show that the conservation of the constraints 
is represented 
by the following Ward identities satisfied by 
$\Lambda ({\bf k},{\bf q},i\omega_{\nu})$ \cite{read98};
\begin{equation}
i \omega_{\nu} \Lambda ({\bf k},{\bf q},i\omega_{\nu})
= i \omega_{\nu} - 
\xi_{{\bf k} + {1 \over 2}{\bf q}} - 
\xi_{{\bf k} - {1 \over 2}{\bf q}} \ .
\label{ward}
\end{equation}
For example, using the Ward identity, one can show that 
$\langle \rho^R ({\bf q}) \rho^R (-{\bf q}) \rangle =
\langle \rho^R ({\bf q}) \rho^L (-{\bf q}) \rangle = 0$
and so on \cite{read98}.

At low temperatures, in the limit of small 
$q$ and $\omega_{\nu}$, the expression for 
$\Gamma ({\bf k},{\bf k'},{\bf q},i\omega_{\nu})$ 
was explicitly found as \cite{read98}
\begin{equation}
\Gamma ({\bf k},{\bf k'},{\bf q},i\omega_{\nu})
= {{\bf q} \cdot {\bf v}_{\bf k} \
{\bf q} \cdot {\bf v}_{\bf k'} \over
\omega^2_{\nu} \chi_0 ({\bf q},i\omega_{\nu})}
- {{\bf q} \wedge {\bf v}_{\bf k} \
{\bf q} \wedge {\bf v}_{\bf k'} \over
q^2 \chi^{\perp}_0 ({\bf q},i\omega_{\nu})} \ ,
\label{smallV}
\end{equation}  
where
\begin{equation}
\chi_0 ({\bf q},i\omega_{\nu}) = -
\int {d^2 k \over (2\pi)^2}
{f(\xi_{{\bf k} + {1 \over 2}{\bf q}})-
f(\xi_{{\bf k} - {1 \over 2}{\bf q}}) \over
\xi_{{\bf k} + {1 \over 2}{\bf q}} - 
\xi_{{\bf k} - {1 \over 2}{\bf q}} - i \omega_{\nu}} \ , 
\label{denres}
\end{equation}
and 
\begin{equation}
\chi^{\perp}_0 ({\bf q},i\omega_{\nu}) = -
{1 \over 2} {\cal N}(0)v^2_F + 
\chi^{\perp}_{0p} ({\bf q},i\omega_{\nu}) 
\label{curres}
\end{equation}
with
\begin{equation}
\chi^{\perp}_{0p}
= - \int {d^2 k \over (2\pi)^2}
\left ( {{\bf {\hat {\bf q}}} 
\wedge {\bf k} \over m^*} \right )^2
{f(\xi_{{\bf k} + {1 \over 2}{\bf q}})-
f(\xi_{{\bf k} - {1 \over 2}{\bf q}}) \over
\xi_{{\bf k} + {1 \over 2}{\bf q}} - 
\xi_{{\bf k} - {1 \over 2}{\bf q}} - i \omega_{\nu}} \ . 
\label{para}
\end{equation}
Here ${\bf v}_{\bf k}$ is the velocity and 
${\cal N}(0)$ the density of states at the Fermi 
level. 
In the long wavelength limit, the exponential factors
in the expression of $\chi^{irr}_{LL}$ can be expanded
as $e^{i {\bf k} \wedge {\bf q}} - 1 \approx 
i {\bf k} \wedge {\bf q}$. Using the scattering vertex
function in the long wavelength and low energy limits,
one finds \cite{stern,read98}
\begin{equation}
\chi^{irr}_{LL} \approx - q^2 {\bar \rho} \  
{{\bar \rho} + m^* \chi^{\perp}_0 ({\bf q},i\omega_{\nu})
\over \chi^{\perp}_0 ({\bf q},i\omega_{\nu})} \ . 
\end{equation}
It leads to the similar result to that of the 
density-density response function in the Chern-Simons 
theory in the long wavelength and low energy limits
(small $|{\bf q}|$ and $\omega$) \cite{read98};
\begin{equation}
\chi^{irr}_{LL} \approx {{\bar \rho}^2 \over
-\chi^*_d - i \omega k_F / (2 \pi q^3)} \ ,
\end{equation} 
where $\chi^*_d$ is the diamagnetic susceptibility
of the Fermi gas with the dispersion $\xi_{\bf k}$.
This also implies that the compressibility of the
system is finite.

\section{q=2$k_F$ density-density response function} 

In this section, we study the $q=2k_F$ density-density
response function in the lowest-Landau-level theory.
The difficulty in finding the finite wavevector response
function comes from the fact that one needs to know 
the scattering vertex function 
$\Gamma ({\bf k},{\bf k'},{\bf q},i\omega_{\nu})$ 
for arbitary ${\bf q}$ and $\omega_{\nu}$.
In principle, if one can find all the eigenvalues and
eigenfunctions of the integral kernel, one can find
the inverse of the integral operator and find the
solution for 
$\Gamma ({\bf k},{\bf k'},{\bf q},i\omega_{\nu})$.
This turns out to be not an easy task for arbitray 
${\bf q}$ and $\omega_{\nu}$.
 
In the small $|{\bf q}|$ limit, the nontrivial part of
the integral kernel becomes 
\begin{equation}
{\tilde V}({\bf k'}-{\bf k}_1) \left . 
{\partial f \over \partial \varepsilon} 
\right |_{\xi_{\bf k}} \ , 
\end{equation} 
which is concentrated at $k=k_F$ at zero temperature.
In this limit, both ${\tilde V}({\bf k}-{\bf k'})$
and the scattering vertex function can be expanded only 
in terms of 
${\rm cos} \ell \theta_{\bf k}$ and 
${\rm sin} \ell \theta_{\bf k}$ ($\ell = 0,1,...$)
because both ${\bf k}$ and ${\bf k'}$ are basically 
restricted on the Fermi surface \cite{read98}.
If one takes only the $\ell = 1$ modes, the 
eigenfunctions are just 
${\bf q} \cdot {\bf v}_{\bf k} / q$ and
${\bf q} \wedge {\bf v}_{\bf k'} / q$.
By finding the correspoding eigenvalues, one gets
the expression of 
$\Gamma ({\bf k},{\bf k'},{\bf q},i\omega_{\nu})$
in Eq.\ref{smallV} \cite{read98}. 

When ${\bf q}$ is not small, the tricks outlined above
do not work. Fortunately, we can find at least
one eigenvector which is correct for all ${\bf q}$
in the small frequency limit. 
This eigenvector can be found by use of the Ward
identity and is given by 
$\xi_{{\bf k}+{1 \over 2}{\bf q}}-
\xi_{{\bf k}-{1 \over 2}{\bf q}}$ with the eigenvalue
tending to zero as $i \omega_{\nu} \rightarrow 0$.
This part of the vertex function can be explicitly
found as
\begin{equation}
\Gamma_1 ({\bf k},{\bf k'},{\bf q},i\omega_{\nu})
= {(\xi_{{\bf k}+{1 \over 2}{\bf q}}-
\xi_{{\bf k}-{1 \over 2}{\bf q}})
(\xi_{{\bf k'}+{1 \over 2}{\bf q}}-
\xi_{{\bf k'}-{1 \over 2}{\bf q}}) \over
\omega^2_{\nu} \chi_0 ({\bf q},i\omega)} \ .
\label{G1}
\end{equation}   
One can also show that 
$\Gamma_1 ({\bf k},{\bf k'},{\bf q},i\omega_{\nu})$
exhausts the Ward identity in Eq.\ref{ward} 
in the following sense. 
Let us consider the following quantity.

\widetext
\top{-2.8cm}
\begin{eqnarray}
{\tilde \Lambda} ({\bf k},{\bf q},i\omega_{\nu})
&=& 1 - \int {d^2 k_1 \over (2\pi)^2}
{f(\xi_{{\bf k}_1 + {1 \over 2}{\bf q}})-
f(\xi_{{\bf k}_1 - {1 \over 2}{\bf q}}) \over
\xi_{{\bf k}_1 + {1 \over 2}{\bf q}} - 
\xi_{{\bf k}_1 - {1 \over 2}{\bf q}} - i \omega_{\nu}}
\ \Gamma_1 ({\bf k}_1,{\bf k},{\bf q},i\omega_{\nu}) \cr
&=& 1 - {\xi_{{\bf k}+{1 \over 2}{\bf q}}-
\xi_{{\bf k}-{1 \over 2}{\bf q}} \over
\omega^2_{\nu} \chi_0 ({\bf q},i\omega)}
\int {d^2 k_1 \over (2\pi)^2}
{f(\xi_{{\bf k}_1 + {1 \over 2}{\bf q}})-
f(\xi_{{\bf k}_1 - {1 \over 2}{\bf q}}) \over
\xi_{{\bf k}_1 + {1 \over 2}{\bf q}} - 
\xi_{{\bf k}_1 - {1 \over 2}{\bf q}} - i \omega_{\nu}}
(\xi_{{\bf k}_1+{1 \over 2}{\bf q}}-
\xi_{{\bf k}_1-{1 \over 2}{\bf q}}) \cr
&=& 1 - {\xi_{{\bf k}+{1 \over 2}{\bf q}}-
\xi_{{\bf k}-{1 \over 2}{\bf q}} \over
\omega^2_{\nu} \chi_0 ({\bf q},i\omega)}
\int {d^2 k_1 \over (2\pi)^2} \left \{
\left [ f(\xi_{{\bf k}_1 + {1 \over 2}{\bf q}})-
f(\xi_{{\bf k}_1 - {1 \over 2}{\bf q}}) \right ] +
i \omega_{\nu}
{f(\xi_{{\bf k}_1 + {1 \over 2}{\bf q}})-
f(\xi_{{\bf k}_1 - {1 \over 2}{\bf q}}) \over
\xi_{{\bf k}_1 + {1 \over 2}{\bf q}} - 
\xi_{{\bf k}_1 - {1 \over 2}{\bf q}} 
- i \omega_{\nu}} \right \} \cr
&=& 1 - {\xi_{{\bf k}+{1 \over 2}{\bf q}}-
\xi_{{\bf k}-{1 \over 2}{\bf q}} \over
i \omega_{\nu}} \ ,
\end{eqnarray} 
where the definition of $\chi_0$ in Eq.\ref{denres}
is used in going from the third to fourth lines.
Notice also that, among two terms in the curly braket 
in the third line, the first term gives zero contribution 
to the ${\bf k}_1$-integral.  
Comparing with the Ward identity in Eq.\ref{ward}, one can see
that ${\tilde \Lambda} ({\bf k},{\bf q},i\omega_{\nu})$ 
is nothing but the one-particle-irreducible vertex
$\Lambda ({\bf k},{\bf q},i\omega_{\nu})$.
Therefore, the full scattering vertex function
can be written as
\begin{equation}
\Gamma ({\bf k},{\bf k'},{\bf q},i\omega_{\nu})
= \Gamma_1 ({\bf k},{\bf k'},{\bf q},i\omega_{\nu})
+ \Gamma_2 ({\bf k},{\bf k'},{\bf q},i\omega_{\nu})
\end{equation} 
such that
\begin{equation}
\int {d^2 k_1 \over (2\pi)^2}
{f(\xi_{{\bf k}_1 + {1 \over 2}{\bf q}})-
f(\xi_{{\bf k}_1 - {1 \over 2}{\bf q}}) \over
\xi_{{\bf k}_1 + {1 \over 2}{\bf q}} - 
\xi_{{\bf k}_1 - {1 \over 2}{\bf q}} - i \omega_{\nu}}
\ \Gamma_2 
({\bf k}_1,{\bf k},{\bf q},i\omega_{\nu}) = 0 \ .
\label{zero}
\end{equation}

Now, for notational convenience, let us rewrite the 
irreducible density-density response function as follows.
\begin{equation}
\chi^{irr}_{LL} ({\bf q},i\omega_{\nu})
= \chi_a ({\bf q},i\omega_{\nu}) + 
\chi_b ({\bf q},i\omega_{\nu}) + 
\chi_c ({\bf q},i\omega_{\nu}) \ , 
\end{equation}
where
\begin{eqnarray}
\chi_a &=&
-\int {d^2 k \over (2\pi)^2}
(e^{i {\bf k} \wedge {\bf q}}-1)
(e^{-i {\bf k} \wedge {\bf q}}-1)
{f(\xi_{{\bf k} + {1 \over 2}{\bf q}})-
f(\xi_{{\bf k} - {1 \over 2}{\bf q}}) \over
\xi_{{\bf k} + {1 \over 2}{\bf q}} - 
\xi_{{\bf k} - {1 \over 2}{\bf q}} - i \omega_{\nu}} \ , \cr
\chi_b &=&
\int {d^2 k \ d^2 k' \over (2\pi)^4} 
(e^{i {\bf k} \wedge {\bf q}}-1)
{f(\xi_{{\bf k} + {1 \over 2}{\bf q}})-
f(\xi_{{\bf k} - {1 \over 2}{\bf q}}) \over
\xi_{{\bf k} + {1 \over 2}{\bf q}} - 
\xi_{{\bf k} - {1 \over 2}{\bf q}} - i \omega_{\nu}}
\Gamma_1 ({\bf k},{\bf k'},{\bf q},i\omega_{\nu})
{f(\xi_{{\bf k'} + {1 \over 2}{\bf q}})-
f(\xi_{{\bf k'} - {1 \over 2}{\bf q}}) \over
\xi_{{\bf k'} + {1 \over 2}{\bf q}} - 
\xi_{{\bf k'} - {1 \over 2}{\bf q}} - i \omega_{\nu}}
(e^{-i {\bf k'} \wedge {\bf q}}-1) \ , \cr
\chi_c &=&
\int {d^2 k \ d^2 k' \over (2\pi)^4} 
(e^{i {\bf k} \wedge {\bf q}}-1)
{f(\xi_{{\bf k} + {1 \over 2}{\bf q}})-
f(\xi_{{\bf k} - {1 \over 2}{\bf q}}) \over
\xi_{{\bf k} + {1 \over 2}{\bf q}} - 
\xi_{{\bf k} - {1 \over 2}{\bf q}} - i \omega_{\nu}}
\Gamma_2 ({\bf k},{\bf k'},{\bf q},i\omega_{\nu})
{f(\xi_{{\bf k'} + {1 \over 2}{\bf q}})-
f(\xi_{{\bf k'} - {1 \over 2}{\bf q}}) \over
\xi_{{\bf k'} + {1 \over 2}{\bf q}} - 
\xi_{{\bf k'} - {1 \over 2}{\bf q}} - i \omega_{\nu}}
(e^{-i {\bf k'} \wedge {\bf q}}-1) \ .
\label{chi}
\end{eqnarray}
\bottom{-2.7cm}
\narrowtext
\noindent
Let us investigate $\chi_a$ and $\chi_b$ first, then
consider $\chi_c$ later.
When the momentum transfer is given by 
${\bf q} = {\bf Q}_0 \equiv 2 k_F {\bf {\hat {\bf q}}}$, 
one can see that ${\bf k} \pm {1 \over 2} {\bf Q}_0$ can be 
rewritten as
\begin{equation}
{\bf k} \pm {1 \over 2} {\bf Q}_0 = 
\pm k_F {\bf {\hat {\bf q}}} + {\bf p} \ . 
\end{equation}    
We can also easily see that 
${\bf k} \wedge {\bf Q}_0 = 
{\bf p} \wedge {\bf Q}_0 = 2k_F {\bf p} \wedge 
{\bf {\hat {\bf q}}}$,
where ${\bf p} \wedge {\bf {\hat {\bf q}}}$ is the 
component of ${\bf p}$ perpendicular to ${\bf {\hat {\bf q}}}$. 
In the low energy limit $|\omega| \ll \varepsilon_F$ (here
$\varepsilon_F \equiv \varepsilon_{|{\bf k}|=k_F}$),
we have ${\bf p} \wedge {\bf {\hat {\bf q}}} \ll k_F$ in
the imaginary part of the analytically-continued 
response functions
$\chi ({\bf q},i\omega_{\nu} \rightarrow \omega + i0^+)$.
In this case, restoring $l^2_B = 
(2/{\tilde \phi})k^{-2}_F$ in the expression,
the exponential factor in $\chi_a$ can be expanded as 
\begin{equation}
e^{i l^2_B {\bf k} \wedge {\bf q}} - 1 = 
e^{i l^2_B 2 k_F {\bf p} \wedge {\bf {\hat {\bf q}}}} - 1 
\approx i l^2_B 2 k_F {\bf p} \wedge {\bf {\hat {\bf q}}} \ .
\end{equation}   
Thus the imaginary part of $\chi_a ({\bf Q}_0,\omega)$ 
in the low energy limit can be written as
\begin{equation}
{\rm Im} \chi_a ({\bf Q}_0,\omega) \approx
2 \left ( {2 \over {\tilde \phi}} \right )^2 
\left ( {m^* \over k_F} \right )^2 \  
{\rm Im} \chi^{\perp}_{0p} ({\bf Q}_0,\omega) \ , 
\end{equation} 
where $\chi^{\perp}_{0p} ({\bf q},i\omega_{\nu})$ is given
by Eq.\ref{para}.
After changing the integral variable from ${\bf k}$ to
${\bf p}$ defined via 
${\bf k} \pm {1 \over 2} {\bf Q}_0 = 
\pm k_F {\bf {\hat {\bf q}}} + {\bf p}$, 
one can take the low frequency
approximation $|\omega| \ll \varepsilon_F$ and
${\bf p} \wedge {\bf {\hat {\bf q}}}, 
{\bf p} \cdot {\bf {\hat {\bf q}}} \ll k_F$ to find
the imaginary part of 
$\chi^{\perp}_{0p} ({\bf Q}_0,i\omega_{\nu})$.
Here we can use 
\begin{eqnarray}
\xi_{\bf k} &=& \xi (|{\bf k}|)
= \xi \left ( |-k_F {\bf {\hat {\bf q}}} + {\bf p}| \right ) \cr
&\approx& \xi \left ( k_F-{\bf p} \cdot {\bf {\hat {\bf q}}}
+ {1\over 2} 
{({\bf p} \wedge {\bf {\hat {\bf q}}})^2 \over k_F} \right ) \cr
&\approx& \xi_{k_F} + \left ( -{\bf p} \cdot {\bf {\hat {\bf q}}}
+ {1\over 2} 
{({\bf p} \wedge {\hat {\bf q}})^2 \over k_F} \right ) \left (
{\partial \xi_{\bf k} \over \partial |{\bf k}|} 
\right )_{|{\bf k}|=k_F} \cr
&=&  -{k_F \over m^*}{\bf p} \cdot {\bf {\hat {\bf q}}}
+ {1\over 2} 
{({\bf p} \wedge {\hat {\bf q}})^2 \over m^*} \ .
\end{eqnarray}
Similarly, we get
\begin{equation}
\xi_{{\bf k}+{\bf q}} 
\approx {k_F \over m^*}{\bf p} \cdot {\bf {\hat {\bf q}}}
+ {1\over 2} 
{({\bf p} \wedge {\hat {\bf q}})^2 \over m^*} \ .
\end{equation}
After finding the imaginary part, the leading order 
behavior of the real part
can be also found by the Kramers-Kronig relation with 
a frequency cutoff that depends on the details of the 
interaction potential. Finally, we get

\widetext
\top{-2.8cm}
\begin{equation}
\chi^{\perp}_{0p} ({\bf Q}_0,\omega) \approx
{m^* \over 2\pi}
\left ( {k_F \over m^*} \right )^2 
\left [ C_1 \left ( 
{2 m^* \varepsilon_F \over k^2_F} \right )^{3/2} 
+ i {1 \over 6} \left ( 
{m^* \omega \over k^2_F} \right )^{3/2} \right ] \ ,
\label{paraLLL} 
\end{equation}
where $C_1$ is a constant that depends on
the details of the single-particle spectrum $\xi_{\bf k}$.
Recall that $\varepsilon_F = \varepsilon_{|{\bf k}|=k_F}$.
In the case of the quadratic band, $\xi_{\bf k} = 
k^2/2m^* - \mu$, $C_1=1/6$ and $2 m^* \varepsilon_F/k^2_F=1$. 
Using Eq.\ref{paraLLL}, the low frequency limit of 
${\rm Im} \chi_a ({\bf Q}_0,\omega)$ 
can be estimated as
\begin{equation}
{\rm Im} \chi_a ({\bf Q}_0,\omega) \approx 
\left ( {2 \over {\tilde \phi}} \right )^2 
{m^* \over 6 \pi}
\left ( {m^* \omega \over k^2_F} \right )^{3/2} \ . 
\end{equation}

Similarly, ${\rm Im} \chi_b ({\bf Q}_0,\omega)$ 
can be evaluated by expanding the exponential factors.
Notice, however, that the expansion in the linear 
order in $i l^2_B 2 k_F {\bf p} \wedge {\bf {\hat {\bf q}}}$
will give zero contribution to the integral because of the
symmetry of the integrand. Thus one has to keep the terms
that are second order in
$i l^2_B 2 k_F {\bf p} \wedge {\bf {\hat {\bf q}}}$.
After some algebra, we obtain the following result
in the low frequency limit.
\begin{equation}
{\rm Im} \chi_b ({\bf Q}_0,\omega) \approx
- 4 \left ( {2 \over {\tilde \phi}} \right )^4
\left ( {m^* \over k_F} \right )^4 {\rm Im} \left \{ 
{[\chi^{\perp}_{0p} ({\bf Q}_0,\omega)]^2 \over
\chi_0 ({\bf Q}_0,\omega)} \right \} \ .  
\end{equation}  
In the low frequency limit, one can also show that
\begin{equation}
\chi_0 ({\bf Q}_0,\omega) \approx
{m^* \over 2\pi}
\left [ D_1 
\left ( {2 m^* \varepsilon_F \over k^2_F} \right )^{1/2}
+ i {1 \over 2} \left ( 
{m^* \omega \over k^2_F} \right )^{1/2} \right ] \ , 
\label{denLLL}
\end{equation}
where $D_1$ is a constant that depends on
the details of the single-particle spectrum $\xi_{\bf k}$.
In the case of the quadratic band with the effective mass
$m^*$, we get $D_1=1$ with $2 m^* \varepsilon_F/k^2_F=1$. 
Using Eq.\ref{paraLLL} and 
Eq.\ref{denLLL}, the leading order contribution
to ${\rm Im} \chi_b ({\bf Q}_0,\omega)$ can be estimated as
\begin{eqnarray}
{\rm Im} \chi_b ({\bf Q}_0,\omega) &\approx&
4 \left ( {2 \over {\tilde \phi}} \right )^4
\left ( {m^* \over k_F} \right )^4 
\left [ {{\rm Re} \chi^{\perp}_{0p} ({\bf Q}_0,\omega) \over
{\rm Re} \chi_0 ({\bf Q}_0,\omega)} \right ]^2
\ {\rm Im} \chi_0 ({\bf Q}_0,\omega) \cr  
&\approx& \left ( {2 \over {\tilde \phi}} \right )^4
{C_1^2 \over D^2_1} 
\left ( {2 m^* \varepsilon_F \over k^2_F} \right )^2 
{m^* \over \pi}
\left ( {m^* \omega \over k^2_F} \right )^{1/2} \ .
\end{eqnarray}

\bottom{-2.7cm}
\narrowtext
Now let us consider $\chi_c ({\bf q},\omega)$. 
In principle, in order to evaluate $\chi_c$ for finite
${\bf q}$, one needs to know the form of 
$\Gamma_2 ({\bf k},{\bf k'},{\bf q},i\omega_{\nu})$ 
for arbitrary ${\bf k}$ and ${\bf k'}$. 
This is a difficult task 
because the form of $\Gamma_2$ cannot be obtained 
from Ward identities. As a result,
$\Gamma_2$ will, in general, depend on some
detailes of the given potential ${\tilde V}({\bf q})$.  
Some progress can be made, however, if the momentum 
transfer ${\bf q}$ is given by 
${\bf Q}_0=2k_F {\bf {\hat {\bf q}}}$.    
It is also worthwhile to notice that, due to the identity
given by Eq.\ref{zero}, 
$\Gamma_2 ({\bf k},{\bf k'},{\bf q},i\omega_{\nu})$
is an odd function of ${\bf k} \wedge {\bf q}$ and 
${\bf k'} \wedge {\bf q}$. 
In fact, the small ${\bf q}$ limit of 
$\Gamma_1 ({\bf k},{\bf k'},{\bf q},i\omega_{\nu})$
becomes the first term in the expression of the small 
${\bf q}$ limit of the full scattering 
vertex in Eq.\ref{smallV}.
Thus the second term in Eq.\ref{smallV} 
can be regarded as the
small ${\bf q}$ limit of $\Gamma_2$. This form, of course,
cannot be easily generalized to the case of arbitrary
${\bf q}$ and $\omega$. 
  
Let us investigate the integral equation for 
$\Gamma ({\bf k},{\bf k'},{\bf Q}_0,i\omega_{\nu})$
given by Eq.\ref{integral} more closely in the small
$\omega_{\nu}$ limit.
In order to make a progress, let us define 
${\bf p}$ and ${\bf p'}$ such that
${\bf k} \pm {1 \over 2} {\bf q} = \pm k_F 
{\bf {\hat {\bf q}}} + {\bf p}$ and 
${\bf k'} \pm {1 \over 2} {\bf q} = \pm k_F 
{\bf {\hat {\bf q}}} + {\bf p'}$.
In the imaginary part of $\chi_c ({\bf Q}_0,\omega)$,
we have ${\bf p} \wedge {\bf {\hat {\bf q}}}, \  
{\bf p'} \wedge {\bf {\hat {\bf q}}} \ll k_F$
in the low frequency limit. This means that
${\bf k}-{\bf k'} = {\bf p}-{\bf p'} \ll k_F$.
In this case, it is reasonable to assume that 
${\tilde V}({\bf k}-{\bf k'})
= {\tilde V}({\bf p}-{\bf p'})$ can be expanded 
in terms of ${\bf p} \cdot {\bf {\hat {\bf q}}}$,
${\bf p'} \cdot {\bf {\hat {\bf q}}}$, 
${\bf p} \wedge {\bf {\hat {\bf q}}}$,
and ${\bf p'} \wedge {\bf {\hat {\bf q}}}$.

One can see that similar consideration applies to 
the ${\bf k}_1$-dependence of the kernel of the 
integral equation in Eq.\ref{integral} 
as far as the frequency
$\omega_{\nu}$ is sufficiently small.
One has to, however, introduce a cutoff in the 
${\bf p}_1$ integral after changing the variable from
${\bf k}_1$ to ${\bf p}_1$. This cutoff depends on the details
of the potential ${\tilde V}({\bf q})$.   
In view of the structure of the integral equation,
it is also reasonble to assume that, in the low
frequency limit, the scattering vertex 
$\Gamma ({\bf k},{\bf k'},{\bf Q}_0,i\omega_{\nu})$ 
can be expanded in terms of 
${\bf p} \cdot {\bf {\hat {\bf q}}}$,
${\bf p'} \cdot {\bf {\hat {\bf q}}}$,
${\bf p} \wedge {\bf {\hat {\bf q}}}$, and  
${\bf p'} \wedge {\bf {\hat {\bf q}}}$. 
In fact, these are the eigenfunctions of the integral
equation in the low frequency limit.
More explicitly, the integral equation in the low
frequency limit may be rewritten as
\widetext
\top{-2.8cm}
\begin{equation}
\Gamma ({\bf p},{\bf p'},{\bf Q}_0,i\omega_{\nu}) = 
{\tilde V}({\bf p'}-{\bf p}) - 
\int' {d^2 p_1 \over (2\pi)^2}
\Gamma ({\bf p},{\bf p}_1,{\bf q},i\omega_{\nu})
{f(\xi_{k_F {\bf {\hat {\bf q}}} + {\bf p}_1})-
f(\xi_{-k_F {\bf {\hat {\bf q}}} + {\bf p}_1}) \over
\xi_{k_F {\bf {\hat {\bf q}}} + {\bf p}_1} - 
\xi_{-k_F {\bf {\hat {\bf q}}} + {\bf p}_1} - 
i \omega_{\nu}}
{\tilde V}({\bf p}_1-{\bf p'}) \ ,
\end{equation}  
where 
\begin{equation}
{\tilde V}({\bf p}-{\bf p'}) =  
V_1  \ {\bf p} \cdot {\bf {\hat {\bf q}}}
\ {\bf p'} \cdot {\bf {\hat {\bf q}}} + 
V_2  \ {\bf p} \wedge {\bf {\hat {\bf q}}} \  
{\bf p'} \wedge {\bf {\hat {\bf q}}} 
\end{equation}
and $\int'$ represents the fact that there is a cutoff in the
${\bf p}_1$-integral. The solution for 
$\Gamma ({\bf p},{\bf p'},{\bf Q}_0,i\omega_{\nu})$
can be written as
\begin{equation}
\Gamma ({\bf p},{\bf p'},{\bf Q}_0,i\omega_{\nu})
= {{\bf p} \cdot {\bf {\hat {\bf q}}}
\ {\bf p'} \cdot {\bf {\hat {\bf q}}} \over 
\lambda_1 ({\bf Q}_0,i\omega_{\nu})} + 
{{\bf p} \wedge {\bf {\hat {\bf q}}} \  
{\bf p'} \wedge {\bf {\hat {\bf q}}} \over
\lambda_2 ({\bf Q}_0,i\omega_{\nu})} \ .
\end{equation}
One can see that the integral equations for the 
first and the second terms are decoupled.  
The first term is nothing but the ${\bf q}={\bf Q}_0$ limit
of $\Gamma_1$ in Eq.\ref{G1}. 
Thus we know exactly what 
$\lambda_1 ({\bf Q}_0,i\omega_{\nu})$ is. 
The second term would correspond to a contribution from 
$\Gamma_2$. The explicit form can be found from the
integral equation as
\begin{equation}
\Gamma_2 ({\bf p},{\bf p'},{\bf Q}_0,i\omega_{\nu})
\approx {{\bf p} \wedge {\bf {\hat {\bf q}}} \  
{\bf p'} \wedge {\bf {\hat {\bf q}}} \over
V^{-1}_2 + (m^*)^2 \chi^{\perp}_{0p}
({\bf Q}_0,i\omega_{\nu})}
\label{G2}
\end{equation}
in the small $\omega$ limit.

Using $\Gamma_2$ obtained above, we now evaluate ${\rm Im} \chi_c$.
In the low frequency limit, the exponential factor can 
be again expanded by assuming
that $l^2_B {\bf k} \wedge {\bf q} = 
l^2_B 2k_F {\bf p} \wedge {\bf {\hat {\bf q}}} \ll 1$
or ${\bf p} \wedge {\bf {\hat {\bf q}}} \ll k_F$. 
Substituting $\Gamma_2$ in Eq.\ref{G2} to the expression 
of $\chi_c$ in Eq.\ref{chi}, we get
\begin{equation}
{\rm Im} \chi_c ({\bf Q}_0,\omega) =
4 \left ( {2 \over {\tilde \phi}} \right )^2
{(m^*)^4 \over k^2_F} \ {\rm Im} \left \{ 
{[\chi^{\perp}_{0p} ({\bf Q}_0,\omega)]^2 \over
V^{-1}_2 + (m^*)^2 
\chi^{\perp}_{0p} ({\bf Q}_0,\omega)}
\right \} \ .
\end{equation}
In the low frequency limit, the leading order 
contribution is given by
\begin{eqnarray}
{\rm Im} \chi_c ({\bf Q}_0,\omega) &\approx&
4 \left ( {2 \over {\tilde \phi}} \right )^2
{(m^*)^4 \over k^2_F} \ 
{[2 V^{-1}_2 + (m^*)^2 \ {\rm Re} \chi^{\perp}_{0p}]
\ {\rm Re} \chi^{\perp}_{0p} \over
(V^{-1}_2 + (m^*)^2 \ {\rm Re} \chi^{\perp}_{0p})^2} \ 
{\rm Im} \chi^{\perp}_{0p} \cr
&\approx&
\left ( {2 \over {\tilde \phi}} \right )^2
{1 + 2 [V_2 (m^* k^2_F / 2 \pi) 
(2 m^* \varepsilon_F / k^2_F)^{3/2} C_1]^{-1} \over
\left \{ 1 + [V_2 (m^* k^2_F / 2 \pi) 
(2 m^* \varepsilon_F / k^2_F)^{3/2} C_1]^{-1} 
\right \}^2 }
\ {m^* \over 3 \pi}
\left ( {m^* \omega \over k^2_F} \right )^{3/2} \ .
\end{eqnarray}
Combining $\chi_a, \chi_b$ and $\chi_c$, we can see that
${\rm Im} \chi_b$ is the leading order contribution in 
${\rm Im} \chi^{irr}_{LL} ({\bf Q}_0,\omega)$. 
Therefore, in the low frequency limit, we have
\begin{equation}
{\rm Im} \chi^{irr}_{LL} \approx 
\left ( {2 \over {\tilde \phi}} \right )^4
{C_1^2 \over D^2_1} 
\left ( {2 m^* \varepsilon_F \over k^2_F} \right )^2 
{m^* \over \pi}
\left ( {m^* \omega \over k^2_F} \right )^{1/2}
\left [
1 + {\cal O} \left ( {m^* \omega \over k^2_F} 
\right ) \right ] \ .
\end{equation} 
In the next section, we will compare this result with
that of the Chern-Simons theory.

\bottom{-2.5cm}
\narrowtext
\section{$Q=2k_F$ density-density response function
in the Chern-Simons theory}
 
The irreducible density-density response function in the 
Chern-Simons theory at arbitrary ${\bf q}$ and 
$\omega$ was previously calculated and 
has the following form \cite{HLR}.
\begin{equation}
\chi_{\rm cs} ({\bf q},\omega) = 
{\chi_0 ({\bf q},\omega) \over 1 - 
\left ( {2 \pi {\tilde \phi} \over q} \right )^2
\chi_0 ({\bf q},\omega) 
\chi^{\perp}_0 ({\bf q},\omega) } \ ,  
\end{equation} 
where $\chi_0 ({\bf q},\omega)$ and 
$\chi^{\perp}_0 ({\bf q},\omega)$ are given by 
Eq.\ref{denres} and Eq.\ref{curres} with the 
quadratic dispersion $\xi_{\bf k} = k^2/2m^* - \mu$.

In the case of the quadratic band, in the low 
frequency limit, it can be shown that
the leading order results are given by
\begin{eqnarray}
\chi_0 ({\bf Q}_0,\omega) &\approx& {m^* \over 2\pi} 
\left [ 1 + i {1 \over 2} \left ( {m^* \omega \over k^2_F} 
\right )^{1/2} \right ] \cr
\chi^{\perp}_0 ({\bf Q}_0,\omega) &\approx& 
{m^* \over 6\pi} \left ( {k_F \over m^*} \right )^2 
\left [ -1 + i {1 \over 2} 
\left ( {m^* \omega \over k^2_F} 
\right )^{3/2} \right ]
\end{eqnarray}
Using the results above, the imaginary part of 
$\chi_{cs} ({\bf Q}_0,\omega)$ in the low frequency
limit can be estimated as
\begin{eqnarray}
{\rm Im} \chi_{\rm cs} ({\bf Q}_0,\omega) &\approx&
{ {\rm Im} \chi_0 \over
\left [ 1 - \left ( {2 \pi {\tilde \phi} \over 2k_F} \right )^2
{\rm Re} \chi_0 \ 
{\rm Re} \chi^{\perp}_0 \right ]^2 } \cr
&\approx&
{{m^* \over 4 \pi} \over \left [ 1 + {1 \over 3} \left (
{{\tilde \phi} \over 2} \right )^2 \right ]^2 } 
\left ( {m^* \omega \over k^2_F} \right )^{1/2} \ . 
\end{eqnarray} 

Notice that the density-density response function in the
Chern-Simon theory, ${\rm Im} \chi_{\rm cs} ({\bf Q}_0,\omega)$,
has the same $\sqrt{\omega}$ dependence as that of
the density-density response function, 
${\rm Im} \chi^{irr}_{LL} ({\bf Q}_0,\omega)$, in the
lowest-Landau-level theory.
However, the coefficients are different. 
In fact, the ratio between them becomes
\begin{equation}
{{\rm Im} \chi^{irr}_{LL} \over
{\rm Im} \chi_{\rm cs} } \approx
4 \left ( {2 \over {\tilde \phi}} \right )^4
\left [ 1 + {1 \over 3} \left (
{{\tilde \phi} \over 2} \right )^2 \right ]^2 
{C_1^2 \over D^2_1} 
\left ( {2 m^* \varepsilon_F \over k^2_F} \right )^2 \ . 
\end{equation}
This ratio is, in general, nonuniversal number which
depends on the details of the interaction potential
${\tilde V}({\bf q})$. 
In order to get some feelings about how small or
large the ratio can be, 
let us take the HF dispersion relation, 
$\xi_{\bf k}$, being quadratic in ${\bf k}$ with the 
effective mass $m^*$. In this case,
and $C_1^2 / D^2_1 = 1/36$ and 
$2 m^* \varepsilon_F / k^2_F=1$.
As a result, we get
\begin{equation}
{{\rm Im} \chi^{irr}_{LL} \over
{\rm Im} \chi_{\rm cs} } \approx
{1 \over 9} \left ( {2 \over {\tilde \phi}} \right )^4
\left [ 1 + {1 \over 3} \left (
{{\tilde \phi} \over 2} \right )^2 \right ]^2 \ . 
\end{equation}
Typical ratios are given by
\begin{eqnarray}
{{\rm Im} \chi^{irr}_{LL} \over
{\rm Im} \chi_{\rm cs} } &\approx& {16 \over 81} \approx 0.198 
\ \ \ \ \ \ {\rm for} \ \ {\nu=1/2} \ , \cr 
{{\rm Im} \chi^{irr}_{LL} \over
{\rm Im} \chi_{\rm cs} } &\approx& {49 \over 1296} \approx 0.038 
\ \ \ {\rm for} \ \ {\nu=1/4} \ ,
\label{ratio}
\end{eqnarray}
and the ratio approaches to $1/81 = 0.012$ as 
${\tilde \phi}$ becomes an infinitely large even number.
Of course, these numbers cannot be taken seriously on the face
value because the HF dispersion is quadratic only for 
small $|{\bf k}| \ll k_F$ and the dispersion relation 
for larger $|{\bf k}|$ affects also these numbers. 
However, it may be suggestive for certain purposes. 
In the next section, we will use these estimations
to discuss the results of the recent phonon drag experiment.
 
\section{discussion on the phonon drag experiment}

Recently, the drag resistivity between two $\nu=1/2$
layers was measured when the layer separation is 
much larger than the magnetic length (or typical
interparticle spacing) in each layer.
In this case, the contribution to the drag resistivity
from the interlayer Coulomb interaction is 
substantially suppressed and the electron-phonon
interaction becomes the dominant source of the 
scattering. Theoretically, the drag resistivity can
be evaluated from \cite{bonsager,cdrag}
\begin{eqnarray}
\rho_{21} &=& {-\hbar^2 \over 8 \pi^2 e^2 n_1 n_2 T}
\int^{\infty}_0 dq \ q^3 \cr
&\times& \int^{\infty} d\omega \left | 
{U_{21}({\bf q},\omega) \over \varepsilon 
({\bf q},\omega)} \right |^2 
{{\rm Im} \chi_1 ({\bf q},\omega)
{\rm Im} \chi_2 ({\bf q},\omega) \over
{\rm sinh}^2(\hbar \omega / 2T)} \ , 
\label{drag}
\end{eqnarray}  
where $\chi_{1,2}$ are the density-density response
functions and $n_{1,2}$ the electron densities 
in layer 1 and 2. $U_{21}$ is the interaction between 
the electrons in different layers and $\varepsilon$
is the interlayer dielectric or screening function. 
The details can be found in Ref.\cite{bonsager} and
Ref.\cite{cdrag}.
If the interaction is dominated by the phonon-mediated
interaction, $U_{21}$ is given by the interlayer 
electron-phonon interaction.
When two layers are identical, the matrix element 
for the drag resistivity invloves 
$[{\rm Im} \chi ({\bf q},\omega)]^2$.  

Let us first discuss the situation of the zero 
applied magnetic field. The phonon-mediated 
drag occurs by transfering momentum ${\bf q}$ of the
phonon to excite the particle-hole continuum
of electrons. At zero temperature, the 
particle-hole continuum ceases to exist if 
the momentum transfer $q$ is larger than $2k_F$
at low frequencies. 
Formally, this means that ${\rm Im}\chi$ becomes
very small if $q > 2k_F$ in the low frequency limit. 
Thus when the phonon energy
$\omega_{\rm ph}=cq$ becomes larger than 
$c (2k_F)$, the scattering is suppressed.
At finite temperature $T$, the typical phonon
energy is set by the temperature scale so that,
if $T > T_0 = c(2k_F)$, the drag resistivity
is substantially suppressed. This is why there
is a maximum of $\rho_{21}$ around $T \approx T_0$
in the measured phonon-mediated drag resistivity.    
 
In the case of the $\nu=1/2$ double-layer system,
the underlying ground state is the composite fermion
Fermi sea. Naively, one may expect 
that the drag resistivity will have a maximum
around $T \approx T_{1/2} > T_0$ where 
$T_{1/2}=c(2k^{\rm cf}_F)$ and 
$k^{\rm cf}_F=\sqrt{2} k^e_F$. 
Here $k^{\rm cf}_F$ is the Fermi wave vector 
of composite fermion system.   
In the experiment, the maximum of the drag
resistivity occurs at a temperture that is 
even smaller than $T_0$ \cite{gramila99}.

In order to invetigate the theoretical
consequence in detail, Bonsager, MacDonald, and 
the author calculated the phonon-mediated 
drag resistivity in the Chern-Simons 
theory \cite{bonsager}.  
In fact, since the electron density-density response
function is not simply proportional to 
the density-density response function of 
composite fermions in the Chern-Simons theory,
it is not obvious whether the naive expectation
is valid even in the Chern-Simons theory.
They found that \cite{bonsager} 

{\bf 1.} If the effective mass of 
composite fermions is sufficiently small 
such that the effective Fermi energy is
substantially larger than $T_{1/2}$, the
naive expectation is more or less correct.
That is, the maximum occurs around $T_{1/2}$.

{\bf 2.} When realistic values of the
effective mass are used, however,
the effective Fermi energy 
is not very small compared to typical temperature
scale we are looking at (such as $T_{1/2}$).
As a result, there are significant finite 
temperature effects leading to the shift
(not necessarily monotonic as a function of 
the effective mass) of the maximum position.

{\bf 3.} In any case, the maximum of the drag resistivity
occurs always at a temperature larger than $T_0$. 
Thus the theory cannot explain the large downward 
shift of the maximum position. 

Now the question is what happens in the 
lowest-Landau-level theory. In order to evaluate
the drag resistivity in the lowest-Landau-level 
theory, one needs to know, in principle, the 
imaginary part of the density-density response 
function for arbitrary frequency and wavevector.
This may be necessary especially because the 
effective Fermi energy is not much larger 
than the temperature scales we are interested in.
Evaluation of the density-density response function
at arbitrary ${\bf q}$ and $\omega$ in the 
lowest-Landau-theory is not an easy task because
one has to solve the scattering vertex function
$\Gamma$ for arbitrary wavevector and frequency.
Also the finite wavevector response 
function would contain nonuniversal 
contribution which may depend on the details of
the given interaction potential 
${\tilde V}({\bf q})$.  
 
In this paper, we evaluated the imaginary part
of the density-density response function of electrons
for $q=2k_F$ in the low frequency limit. 
Thus the full evaluation of the drag resistivity
is beyond the scope of the present paper.
However, based on what we got in the last section,
we may be able to speculate what might happen
to the drag resistivity within the lowest-Landau-level
theory. We found in the last section that the 
imaginary parts of the density-density response 
functions in the Chern-Simons and lowest-Landau-level
theory (${\rm Im} \chi_{\rm cs}$ and 
${\rm Im} \chi^{irr}_{LL}$) have the same 
$\sqrt{\omega}$ dependence in 
frequency at $q=2k_F$ in the low frequency limit.
It may appear that this is a disappointing result
because the lowest-Landau-theory does not provide
qualitatively different results.

However, we also learned that the ratio between 
${\rm Im} \chi^{irr}_{LL}$ and 
${\rm Im} \chi_{\rm cs}$ can be quite 
different from unity. For the sake of the order of magnitude 
estimation, if we use a quadratic approximation 
for $\xi_{\bf k}$ in the lowest-Landau-level, 
${\rm Im} \chi^{irr}_{LL}/{\rm Im} \chi_{\rm cs} 
\approx 0.2$ for $\nu=1/2$ as shown in Eq.\ref{ratio}.
In the expression of the drag resistivity in Eq.\ref{drag},
the imaginary part of the density-density response
enters as $[{\rm Im} \chi]^2$.
If the behavior of 
${\rm Im} \chi^{irr}_{LL}/{\rm Im} \chi_{\rm cs}$
for larger frequencies is similar to that of
the low frequency limit, the contribution from
the scattering events with $q$ near $2k_F$ to the drag 
resistivity in the lowest-Landau-level theory
would be $\sim 1/25$ times smaller than that of the 
Chern-Simons theory.
This implies that the large wavevector scattering
may be quantitatively more suppressed in the 
lowest-Landau-level theory compared to the case 
of the Chern-Simons theory.
      
In fact, according to the experimental data, the 
smallest drag resistivity in the measured temperature
range is only $\sim 20$ times smaller than the 
maximum value. Thus, if the drag resistivity near 
$T_{1/2} = c (2k_F)$ is $\sim 1/25$ times smaller
than the expected value, it may appear that the
maximum occurs at a much smaller temperature 
scale, $T^{\rm max}_{1/2} < T_{1/2}$.

The discussion above is based on the density-density
response function at $q=2k_F$ in the low frequency
limit. Thus, at this stage, the discussion above 
is only suggestive and not the satisfactory explanation.
It appears that numerical calculation of the 
response function is necessary to understand fully
the phonon drag experiment.

\section{summary}

In this paper, we consider finite $q$ density-density
response function of the compressible Fermi-liquid-like states
in the lowest Landau level. In particular, we evaluated
the $q=2k_F$ density-density response function in the low 
energy limit within the lowest-Landau-level formalism.
We compare this result with the prediction of the Chern-Simons
theory. We found that the density-density response functions
in both cases are proportional to $\sqrt{\omega}$ in the 
low frequency limit. However, the coefficients can be
quite different and the ratio between them 
is a nonuniversal number. 
We found that the response function at $q=2k_F$ in 
the lowest-Landau-level theory can be substantially 
smaller than that of the Chern-Simons theory. 

Using these results, we speculate that the lowest-Landau-level
theory may explain the suppression 
of the $q=2k_F$ scattering seemingly observed in the 
experiment on the phonon-mediated drag
resistivity in the double layer $\nu=1/2$ system.  
However, the satisfactory understanding of the phonon 
drag experiment may require numerical evaluation of the
density-density response function.

Finally, we would like to mention various other 
possibilities which were not considered in this
paper.    
The lowest-Landau-level theory in the 
present form through the time-dependent HF 
approximation may turn out to be inadequate for
describing the large $q$ or short distance 
behaviors of density-density response of the 
composite fermions.
This is because the picture of the
correlations between the electrons and vortices
itself may break down at short distances 
due to the fact that we are looking at the 
length scale comparable to the distance 
between the electrons and vortices \cite{many}.
The theory, if it exists,  which describes this 
situation and the crossover or transition from 
the present lowest-Landau-level theory to this regime
may provide better understanding of the finite 
$q$ response of the system.
In any case, our work should serve as a useful
starting point for understanding more microscopic
picture of the compressible states in the lowest      
Landau level.   

\acknowledgements

We thank Tom Gramila, Ilya Gruzberg, Hae-Young Kee, 
Dung-Hai Lee, Allan MacDonald, 
Ganpathy Murthy, R. Shankar, and especially 
Nick Read for many helpful discussions.
We also thank Aspen Center for Physics for
its hospitality, where
some parts of this work were performed.
This work was supported by the NSF grant 
No. DMR-9983783 and the Alfred P.
Sloan foundation.

\end{multicols}

\end{document}